\documentclass[aps,twocolumn,prx,longbibliography,floatfix,superscriptaddress,amsmath,amssymb,footnoteinbib,10pt]{revtex4}

\usepackage{graphicx}
\usepackage{epstopdf}
\usepackage{xcolor}
\usepackage{bm}

\usepackage[colorlinks, linkcolor=red,citecolor=blue,urlcolor=blue]{hyperref}
\usepackage[all]{hypcap} 

\begin{document}
\bibliographystyle{apsrev}


\title{Multiple topological transitions driven by the interplay of normal scattering and Andreev scattering

}


\author{A.~A. Kopasov}
\affiliation{Institute for Physics of Microstructures, Russian Academy of Sciences, 603950 Nizhny Novgorod, GSP-105, Russia}
\author{A.~S.~Mel'nikov}
\affiliation{Institute for Physics of Microstructures, Russian Academy of Sciences, 603950 Nizhny Novgorod, GSP-105, Russia}
\affiliation{Sirius University of Science and Technology, 1 Olympic Ave, 354340, Sochi, Russia}


\date{\today}

\begin{abstract}
The effect of multiple topological transitions for electron-hole excitations is discovered
in full shell proximitized semiconducting nanowires with trapped superconducting vortices, recently shown to be
a promising platform for the realization of Majorana states at moderate longitudinal magnetic fields.
The mechanism of such multiple transitions is uncovered and explained to be governed by
the interplay of normal and Andreev reflections from the shell and by the textured spin-orbit coupling inside the semiconductor.
The extensive analysis of emerging propagating and Majorana-type evanescent quasiparticle modes in such Andreev waveguides is performed. Experimentally, these modes reveal themselves in peculiarities of charge and spin-polarized heat transport.
\end{abstract}

\pacs{74.20.-z, 74.20.De, 74.25.-q, 74.25.Dw}

\maketitle

\section{Introduction}
\begin{figure}
\centering
\includegraphics[scale = 0.48]{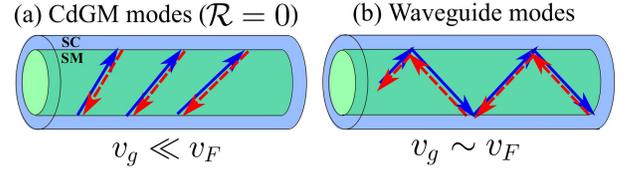}
\caption{Electron (solid) and hole (dashed) trajectories inside the semiconducting core for dominating Andreev (a) and normal (b)  reflection from the wire coating.}
\label{Fig:CdGM_waveguide}
\end{figure}

Recent experimental activity in the studies of Majorana nanowires~\cite{Marcus_arXiv_2018} has revived the interest to the
vortex states in superconducting mesoscopic hybrid systems in the context of their possible applications
for topologically protected quantum computations~\cite{Kitaev,Nayak,Alicea1,Aasen,Alicea2,Elliot,Aguado}. In particular,
it has been suggested~\cite{Lutchyn1} that in semiconducting nanowires fully covered by a superconducting shell the vortices entering the hybrid nanowire in rather low magnetic fields $ H \sim 0.1 $~T along the wire axis $z$
can drive the system into the topological phase even at moderate Zeeman splitting, unlike the original proposal~\cite{LutchynOr,OregOr}. The essential requirement for realization of this proposal is that the vortex entry
should cause the inversion of the energy branches $E_i (k_z)$ of quasiparticles vs. the momentum $k_z$ along the wire similarly to the band inversion in topological insulators~\cite{Volkov, Bernevig} and $^3$He~\cite{Makhlin, Silaev_JLTP} (here $i$ labels spin-dependent transverse modes).
The minigap in the spectrum in this case appears only due to a finite spin-orbit coupling determined by the radial electric field in the nanowire.
Clearly, the desired branch inversion do not take place in the limit of strong proximity effect when the electrons at the semiconductor/superconductor (SM/SC) interface almost completely convert into holes propagating in opposite direction due to the Andreev reflection phenomenon~\cite{Andreev} [see Fig.~\ref{Fig:CdGM_waveguide}(a)].
Indeed, similarly to the case of a vortex core in bulk s-wave superconductors the dominating Andreev reflection should cause the formation of the so -- called Caroli - de Gennes - Matricon (CdGM) minigap~\cite{CdGM,Bardeen} even in the absence of the spin - orbit interaction: $\omega_0 = |\Delta|/k_{\perp}^c R_c$, where $|\Delta|$ is the primary SC gap, $k_{\perp}^c=\sqrt{(k_F^c)^2-k_z^2}$, $k_F^c$ is the Fermi momentum inside the core, and $R_c$ is the core radius. It is this $k_z$ dependence of the minigap which prevents the Fermi level crossings needed for the above topological transitions.
However, the above reasoning completely disregards an important difference between the quasiparticle modes in the vortex core in the bulk superconductor and in the hybrid mesoscopic nanowire: in the latter case one can hardly get the limit of full Andreev reflection at the shell. In addition to this electron -- hole conversion we need to take account of the normal reflection caused by
(i) the jump in the effective masses for the electrons in the SM core $m^*$ and the shell $m_s$, (ii) the jump in the confining potential, and (iii) the normal reflection of quasiparticles from the external SC shell surface.

In this paper we evidently demonstrate the effect of the interplay of the normal and Andreev reflection on the formation of multiple topological transitions in full shell nanowires and provide an extensive analysis of the emerging quasiparticle modes.
Before proceeding let us qualitatively elucidate the main features of this effect on the basis of quasiclassical theory.
Indeed, the interplay of the normal and Andreev reflection processes in hybrid nanowires can be well described within the
quasiclassical theory $k_F^c\xi \gg 1$ using the approach developed previously in Refs.~\cite{Kopnin,Kopnin_PRB_2007,Mel'nikov} for a vortex trapped in a superconducting mesoscopic disk. Here $\xi$ is the superconducting coherence length.
Neglecting the spin-orbit coupling for simplicity we derive the following expression
for the low-energy branches in a singly quantized vortex~\cite{n=1_footnote} (see Appendix~\ref{equation_derivation} for the derivation):
\begin{eqnarray}\label{CdGM_spectrum}
\mathcal{E}_{\mu, k_z}^{\uparrow\downarrow} = \mu(\omega_0 + \omega_H/2) \pm V_Z + \delta \cos\left(\theta_{\mu, k_z}\right) \ ,
\end{eqnarray}
with the second $\omega_H = \hbar eH/m^* c$ and the third $V_Z = g\beta H/2$ terms caused by the orbital~\cite{Hansen} and the Zeeman effect, respectively, while
the last oscillating term with the amplitude $\delta = 2|\Delta|\mathcal{R}(k_z)$ and phase $\theta_{\mu, k_z} = 2k_{\perp}^cR_c - \pi\mu - \pi/2$ is determined by above mentioned normal scattering processes at the SM/SC boundary. Here $\mu$ is half an odd integer, $\hbar$ is the reduced Planck constant, $e > 0$ is an electric charge, $c$ is the speed of light, $g$ is the effective Land\'{e} g-factor in the core, $\beta$ is the Bohr magneton, $\mathcal{R}^2(k_z) \ll 1$ is the probability of the normal scattering at the SM/SC interface, and $\uparrow$ ($\downarrow$) corresponds to the positive (negative) spin projection along the direction of the magnetic field.
Changing the balance between the Andreev and normal reflection $\mathcal{R}$ one can strongly change the hard gap in the quasiparticle spectrum and completely close it when $\delta \sim \omega_0$.
The number of resulting Fermi level crossings (i.e., solutions of the equation $\mathcal{E}_{\mu,k_z} = 0$) depends on the amplitude ratio $\delta /\omega_0=2\mathcal{R}k_{\perp}^c R_c$ and
the number of normal state transverse modes $k_{F}^cR_c$.
Additional fine tuning of the zero-energy modes can be done via the orbital and the Zeeman shifts, cf.~\cite{Makhlin}.
Each of these crossing points gives us an electron -- hole mode at the Fermi level propagating along the Andreev waveguide with the noticeable group velocity $v_g$ of the order of the Fermi velocity $v_F^c$ in the SM core.
Thus, full shell proximitized nanowires provide
a unique possibility to study the crossover between the CdGM states with a vanishing group velocity $v_g\ll v_F^c$ and the discrete Andreev waveguide modes (see Fig.~\ref{Fig:CdGM_waveguide}).

Switching on a finite spin-orbit coupling with the radial texture (introduced similarly to the Ref.~\cite{Lutchyn1})
$$
\hat{H}_{so} = (\alpha/\hbar) \mathbf{e}_{r}\left[\hat{\mathbf{\sigma}},(\mathbf{p} + eA_{\varphi}\mathbf{e}_{\varphi})\right]
$$
results in the hybridization between the states with the neighboring angular momenta
$\mu$ and $\mu+1$. Here $\alpha$ is the Rashba velocity, $\hat{\mathbf{\sigma}}$ is the vector of Pauli matrices acting in the spin space, $\mathbf{p} = -i\hbar\nabla$, $\mathbf{A}$ is the vector potential, $\mathbf{e}_r$ and $\mathbf{e}_{\varphi}$ are the unit vectors in the cylindrical coordinate system. Being most efficient close to the level crossings this hybridization should be particularly important for such crossings at the Fermi level resulting in the opening of the minigaps for the interacting spin partners.

The main goal of our work is to demonstrate that the above series of the waveguide modes gives us the energy branch inversions needed for the formation of a set of topologically nontrivial spin-polarized modes in the Andreev waveguide. Each waveguide mode from this series should also produce the Majorana-type states bound to the nanowire ends. The solution of this problem, in turn,  may provide a way to optimize the topological protection of the Majorana states in multi-channel full shell nanowires even for rather small values of the radial Rashba field~\cite{Woods1}.

The paper is organized as follows. In Sec.~\ref{basic_equations} we introduce the model and the basic equations which are numerically solved to calculate the excitation spectra in multi-channel proximitized nanowires neglecting spin-dependent interactions. In Sec.~\ref{results_and_discussion} the results of our numerical analysis are presented and discussed in context of recent experimental progress on full shell InAs/Al nanowires. In Sec.~\ref{spin_related_effects} we uncover the role of the textured spin-orbit coupling in the formation of multiple topological transitions and show that such transitions are accompanied by the presence of low-energy spin polarized propagating states. Finally, the results are summarized in Sec.~\ref{summary}.

\section{Basic equations}\label{basic_equations}
We support our qualitative consideration of the energy branch inversion by performing numerical calculations of the excitation spectra in full shell wires taking into account both the normal and Andreev scattering and neglecting at the first stage spin-dependent interactions. Note that in further analysis we go beyond the quasiclassical approximation inside the SM core which is of particular importance for understanding the extreme quantum limit $k_F^c\xi\sim 1$ in realistic SM nanowires. Our analysis is based on the Bogoliubov - de Gennes (BdG) equations for the hybrid structure:
\begin{subequations}
\label{BdG_equations}
\begin{align}
\check{\mathcal{H}}\Psi(\mathbf{r}) = E\Psi(\mathbf{r}) \ ,\\
\label{BdG_hamiltonian}
\check{\mathcal{H}} = \check{\tau}_z\left[\boldsymbol{P}\frac{1}{2m(\mathbf{r})}\boldsymbol{P} + U(\mathbf{r})\right] +
\check{\tau}_x{\rm Re}\Delta(\mathbf{r})-\check{\tau}_y{\rm Im}\Delta(\mathbf{r}) \ .
\end{align}
\end{subequations}
Here $\boldsymbol{P} = \left(\mathbf{p} + \check{\tau}_ze\mathbf{A}/c\right)$,  $m(\mathbf{r})$ is the spatially varying effective mass, $U(\mathbf{r})$ is the confining potential, $\check{\tau}_i$ ($i = x,y,z$) are the Pauli matrices acting in the electron~-~hole space, and the superconducting order parameter $\Delta(\mathbf{r})$ is nonzero only in the shell.
For simplicity, we consider the hybrid nanowire of a circular cross-section~\cite{Luo} and use cylindrical coordinates $\mathbf{r} = (r,\varphi,z)$. Neglecting the order parameter inhomogeneity in the shell, we suggest a simplified model describing the magnetic flux dependence of the superconducting gap taking account of the entries of multiquanta vortices~\cite{Self_cons_Delta_footnote}:
\begin{equation}\label{Delta_profile}
\Delta = \Delta_0\left[1 - \gamma\left(\Phi/\Phi_0 - n\right)^2\right]e^{-in\varphi} \ ,
\end{equation}
where $\Delta_0$ is the superconducting gap in the shell at $H = 0$, $n(\Phi) = \lfloor 1/2 + \Phi/\Phi_0 \rfloor$, $\lfloor x \rfloor$ stands for the floor function, $\gamma \sim \xi^2/R_c^2$ for the shell thickness $d_s\ll R_c$, $\Phi = \pi H R_c^2$ and $\Phi_0 = \pi\hbar c/e$ are the magnetic flux through the wire and the flux quantum, respectively. The effects of band bending on the formation of the waveguide modes are modeled by trial
confining potential profiles in the core: (i) a flat potential profile and (ii) a parabolic radial profile~\cite{Self-consist_U_footnote}
\begin{equation}\label{Confining_profile}
U(r) = U_0 - \left(U_0 + E_F^sm^*/m_s\eta^2\right)\left(r/R_c\right)^2 \ ,
\end{equation}
describing the presence of an insulating region in the center of the SM core (when the Fermi level lies below the conduction band edge). Here $E_F^s$ is the Fermi energy in the shell and $\eta$ is the ratio of the Fermi velocities in the shell and the core at $r = R_c$. Radial symmetry of the problem with the chosen gauge of the vector potential $A_{\varphi} = Hr/2$, $A_r = A_z = 0$ allows one to seek the solution of Eqs.~\eqref{BdG_equations} in the form:
\begin{equation}\label{state_ansatz}
\Psi(\mathbf{r}) = e^{ik_zz + i\mu \varphi - i\check{\tau}_zn\varphi/2}\Psi_{\mu, k_z}(r)
\end{equation}
with integer (half an odd integer) $\mu$ for even (odd) $n$ values. Substitution of Eq.~(\ref{state_ansatz}) into Eqs.~(\ref{BdG_equations}) inside the SM core yields the following eigenvalue problem:
\begin{subequations}
\label{eigenvalue_problem}
\begin{align}
\check{\mathcal{L}}\Psi_{\mu, k_z}(r) &= E \Psi_{\mu, k_z}(r) \ ,\\
\nonumber
\check{\mathcal{L}} =  \check{\tau}_z\frac{\hbar^2}{2m^*}&\biggl\{-\frac{1}{r}\frac{d}{dr}\left(r\frac{d}{dr}\right) + \\
&\left[\frac{\mu}{r} + \tau_z\frac{m^*}{\hbar}v(r)\right]^2 +  k_z^2 \biggl\} + \check{\tau}_zU(r) \ .
\end{align}
\end{subequations}
Here $v(r) = (\hbar/2m^*r)(\pi H r^2/\Phi_0 - n)$. We calculate the excitation spectra in full shell wires by using the exact general solutions of Eqs.~(\ref{eigenvalue_problem}) and the appropriate boundary conditions for the quasiparticle wave function at the SM/SC interface which take into account all the aforementioned mechanisms of normal scattering (the derivation of the general solutions and the boundary conditions are shown in Appendices~\ref{exact_solutions}~and~\ref{boundary_conditions_derivation}, respectively). In all our numerical calculations we take a typical value of $\Delta_0/E_F^s = 0.01$ and we also choose the effective mass ratio consistent with the properties of InAs/Al nanowires~\cite{Vurgaftman}: $m^*/m_s = 0.026$.

\section{Numerical simulations}\label{results_and_discussion}
\begin{figure}
\centering
\includegraphics[scale = 0.7]{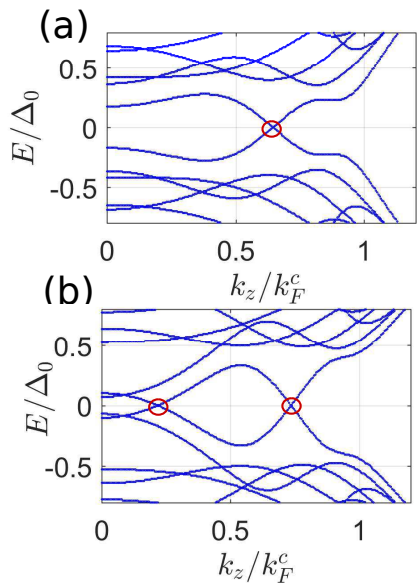}
\caption{Quasiparticle spectra right after the vortex entry ($\Phi = 0.5\Phi_0$) for a flat potential profile, $R_c = 0.7\xi$ and $\gamma = 0.1$. We choose $\eta = 0.75$, $d_s = R_c$ for (a) and $\eta = 0.72$, $d_s = 0.8R_c$ for (b). Here $\xi = \hbar v_F^s/\Delta_0$ and $\gamma$ is defined by Eq.~(\ref{Delta_profile}). Red cirlces show the Fermi level crossings.}
\label{Fig:typical_spectra}
\end{figure}
We proceed with the discussion of the results of numerical simulations.
Typical quasiparticle spectra just after the vortex entry are presented in Fig.~\ref{Fig:typical_spectra}. These plots clearly show that in a semi-quantitative agreement with Eq.~(\ref{CdGM_spectrum}) the interplay between the normal and Andreev reflection at the SM/SC interface in realistic multi-channel nanowires indeed causes multiple branch inversions due to several emerging Fermi level crossings for oscillating sub-gap levels.

\begin{figure}
\centering
\includegraphics[scale = 0.75]{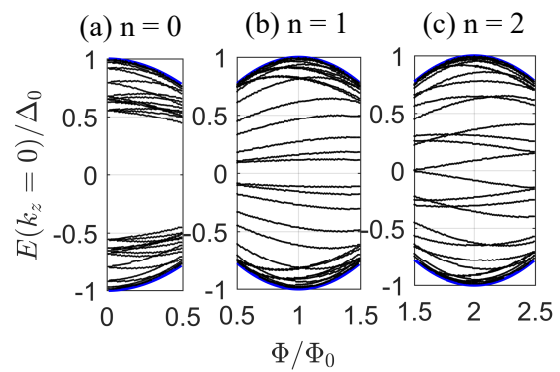}
\caption{Typical magnetic flux dependencies of the levels with $k_z = 0$ for a flat potential profile, $n = 0$ (a), $n = 1$ (b), $n = 2$ (c), $\eta = 0.8$, $R_c = \xi$, $\gamma = 0.9$, and $d_s\gtrsim \xi$. (Blue) solid line shows corresponding $|\Delta(\Phi)|$ profile defined by Eq.~(\ref{Delta_profile}).}
\label{Fig:smiles}
\end{figure}

We show typical magnetic flux dependencies of the sub-gap levels with $k_z = 0$ in Fig.~\ref{Fig:smiles}. Blue solid line shows the corresponding $|\Delta(\Phi)|$ profile defined by Eq.~(\ref{Delta_profile}). We clearly see from Fig.~\ref{Fig:smiles} that $E(\Phi)$ dependencies possess right- or leftward skewness with the respect to lobe centers at $\Phi = n\Phi_0$. This skewness originates from the level shifts $\mu\omega_H/2$ caused by the orbital mechanism and its correct shape requires proper accounting of the non-quasiclassical effects. Obviously, the Zeeman effect which also gives the contribution linear in $H$ should also contribute to the above skewness of the curves as well as to their spin-dependent splitting. We stress that experimentally observed skewed magnetic field dependencies of the sub-gap levels in full shell InAs/As nanowires~\cite{Marcus_private} are in a semi-quantitative agreement with our numerical results, thus, justifying the validity of our model.

\begin{figure}
\centering
\includegraphics[scale = 0.65]{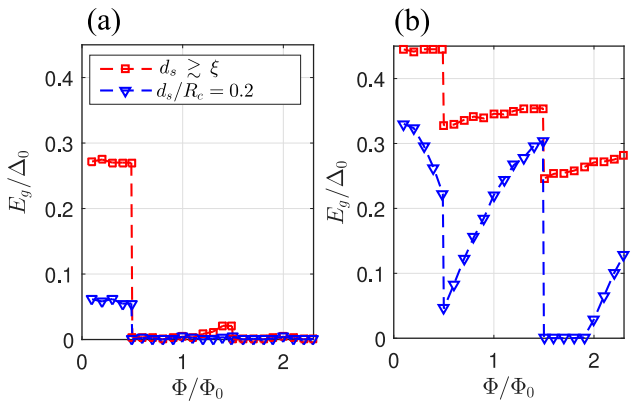}
\caption{Magnetic flux dependencies of the hard gap $E_g$ for $R_c = 0.8\xi$, $\gamma = 0.1$ and different $d_s$. We choose a flat potential profile with $\eta = 0.6$ for (a). We use the potential profile~(\ref{Confining_profile}) with $U_0 = \Delta_0$ and $\eta = 0.5$ for (b).}
\label{Fig:hard_gaps}
\end{figure}

The effects of band bending on the formation of energy branch inversions and corresponding propagating quasiparticle modes in Andreev waveguides are illustrated in Fig.~\ref{Fig:hard_gaps}. These plots clearly show reentrant magnetic flux dependencies of the hard gap due to the level shifts by the magnetic field and the stepwise reduction of the hard gap at vortex entries. Comparing the panels (a) and (b) in Fig.~\ref{Fig:hard_gaps} we see that this reentrant behavior is more pronounced in the case of realistic parabolic $U(r)$ profile modeling the quasiparticle depletion region in the center of the SM core. Certainly, to get the energy branch inversions needed for the topological transitions one should avoid the parameter regions resulting in the hard gap reopenings.
The reentrant magnetic field behavior of the hard gap appears to be in a good agreement with recent measurements of the parity effect in the charge transport through the wire placed in an applied magnetic field and Coulomb blockade conditions ~\cite{Marcus_private}. The magnetic field behavior of the hard gap shown in Fig.~\ref{Fig:hard_gaps}(b) suggests a natural explanation for the observation of a sudden switch of the current periodicity from 2$e$ to $e$ at the vortex entry followed by a smooth transition back to the $2e$ with the further increase in the magnetic flux within a given Little-Parks lobe~\cite{LP62, LP}. Indeed, considering a single vortex state as an example and taking typical $ E_g = 0.25 \Delta_0$ [see Fig.~\ref{Fig:hard_gaps}(b)] in the case of Al shell with $\Delta_0 \approx 200$~ $\mu$eV one can estimate
the appropriate temperature range for the observation of the above odd/even effect $T\lesssim T^* = E_g/\ln(N_{\text{eff}})\lesssim  0.6$~K~\cite{Averin,Tuominen}. It should be noted that our simulations of $E_g(\Phi)$ dependencies are performed without spin-orbit interaction and, thus, do not take into account the presence of possible Majorana zero modes in the system.

To sum up this part of our results, we clearly observe the formation of multiple Fermi level crossings corresponding to the waveguide modes discussed in the introduction. This observation provides us a starting point for further analysis of the spin -- dependent effects needed for a full description of the topological transition mechanism.

\section{Multiple topological transitions and spin-polarized propagating quasiparticle states}\label{spin_related_effects}
Let us proceed with the study of the spin-dependent effects.
Treating the spin-orbit coupling perturbatively, one should take the following expression for the radial part of the quasiparticle wave function:
\begin{equation}\label{wf_spin_degenerate}
\Psi_{\mu,k_z}(r) = C_{\mu,k_z}^{\uparrow}\begin{bmatrix}\
u_{\mu,k_z}(r)\\ v_{\mu,k_z}(r)\\0\\0\end{bmatrix} + C_{\mu,k_z}^{\downarrow}\begin{bmatrix}0\\0 \\ u_{\mu+1,k_z}(r)\\v_{\mu+1,k_z}(r)\end{bmatrix} \ ,
\end{equation}
where $C_{\mu,k_z}^{\uparrow}$ and  $C_{\mu,k_z}^{\downarrow}$ are constants, $u_{\mu,k_z}(r)$ and $v_{\mu,k_z}(r)$ are the electron and hole components of the unperturbed wave function. Substituting Eq.~(\ref{wf_spin_degenerate}) into the full BdG equations with the spin-dependent terms (see Appendix~\ref{spin_orbit_splitting_derivation} for details) and neglecting the spin-orbit induced diagonal corrections to the CdGM levels, we get the following effective model:
\begin{eqnarray}\label{two_level_system}
\left[\mathcal{E}^+_{\mu} + \left(\mathcal{E}^-_{\mu} + V_Z\right)\hat{\sigma}_z + \tilde{\alpha}k_z\hat{\sigma}_y\right]\psi_{\mu,k_z} = E\psi_{\mu,k_z} \ .
\end{eqnarray}
Here $\mathcal{E}^\pm_{\mu,k_z} = [\mathcal{E}_{\mu,k_z}\pm\mathcal{E}_{\mu + 1,k_z}]/2$, $\mathcal{E}_{\mu,k_z}$ is the unperturbed quasiparticle spectrum, $\psi_{\mu,k_z} = [C_{\mu,k_z}^{\uparrow},C_{\mu,k_z}^{\downarrow}]^{{\rm T}}$, $\tilde{\alpha}(k_z) = \alpha M_{\mu,k_z}$, the overlap integral
\begin{eqnarray*}
M_{\mu,k_z} = \int_0^{R_c + d_s} dr \  r [u_{\mu + 1,k_z}(r)u_{\mu, k_z}(r) \\
 - v_{\mu + 1,k_z}(r)v_{\mu, k_z}(r)] \ .
\end{eqnarray*}
Estimating the latter as $M_{\mu,k_z} \sim 1/k_{\perp}^cR_c$ in the limit $k_F^c\xi \sim 1$, we get the equation describing the splitted energy levels
\begin{equation}\label{two_level_spectrum}
E_{\mu,k_z}^{\pm}= \mathcal{E}^+_{\mu,k_z} \pm \sqrt{\left(\mathcal{E}^-_{\mu,k_z} + V_Z\right)^2 + \left(\alpha k_z/k_{\perp}^cR_c\right)^2} \ .
\end{equation}
One can see that the effective model~(\ref{two_level_system}) for $\mu=-1/2$ linearized with respect to $k_z$ near the Fermi crossing points $k_i$ 
\begin{equation}\label{linearized_spectrum}
\mathcal{E}^-_{-1/2}(k_z) \approx v_g^i(k_z-k_i) 
\end{equation}
can be mapped to the one describing a one-dimensional spinless p-wave superconductor~\cite{LutchynOr,OregOr,Oppen_lectures}. Here $v_g^i$ is the Fermi velocity corresponding to the $i$-th waveguide quasiparticle mode.

To define the topological invariants for our model Hamiltonian describing the low-energy states~(\ref{two_level_system}) for $\mu = -1/2$ together with Eq.~(\ref{linearized_spectrum}) we follow the Ref.~\cite{Qi} (see also references therein) and consider the general form of the Hamiltonian $H(k_z) = \mathbf{d}(k_z)\bm{\sigma}$ where in our case the vector $\mathbf{d}$ lies in the $zy$-plane. To elucidate the topological properties of this system we calculate the winding number $W$ of the unit vector $\mathbf{d}/|\mathbf{d}|$ while $k_z$ changes from $-k_F^c$ to $+k_F^c$
\begin{eqnarray}
 W = \frac{1}{2\pi}\int_{-k_F^c}^{k_F^c}d\theta(k_z) \ ,\\ 
 \nonumber
 \theta(k_z) = \arctan\left[d_y(k_z)/d_z(k_z)\right] \ .
\end{eqnarray}
Further topological classification of the quasiparticle states depends on the energy interval under consideration which gives us different forms of the effective Hamiltonian. Starting from the low-energy limit ($|E|\ll \delta$) we find the set of points $\pm k_i$ where the oscillation spectral branch $\mathcal{E}_{-1/2}^-(k_z)$ crosses the Fermi level. One can view these points as multiple Fermi surfaces. Neglecting the coupling between these 1D Fermi surfaces, one can find the topological index for each Fermi surface to be equal $\pm 1$. Correspondingly, each pair of the Fermi level crossings at $\pm k_i$ produces zero-energy evanescent modes at the nanowire edges with the characteristic decay length (see Appendix~\ref{spin_orbit_splitting_derivation} for details)
\begin{equation}\label{xi_loc}
\xi_{loc} \sim R_c\frac{v_g^ik_F^c }{\alpha k_i} \ .
\end{equation}
Taking $v_g^i \sim v_F^c$ and $k_i \sim k_F^c$ in Eq.~\eqref{xi_loc}, we get $\xi_{loc}\gg R_c$ in the case of weak spin-orbit coupling $\alpha \ll v_F^c$.
Thus, we obtain a series of Majorana partner states localized near the wire edges and a number of these Majorana pairs equals to the number of Fermi level crossings.
The number of Fermi level crossings can be tuned by changing the coefficient $\mathcal{R}$ of the normal reflection. Increasing this coefficient, e.g., by shrinking the shell thickness $d_s$ we increase the amplitude of level oscillations. Each time the oscillating curve $\mathcal{E}_{-1/2}^-(k_z)$ touches the Fermi level a new pair of Fermi level crossing points appears. Near each crossing point the spectrum splits due to the spin-orbit coupling and a pair of Majorana states is produced. The corresponding level splitting is defined by the expression $\alpha k_i/R_c\sqrt{(k_F^c)^2 - k_i^2} \sim \alpha/R_c$. Taking $\alpha \approx 50$~meV~\AA~\cite{Woods1} and $R_c\approx 70$~nm for a typical nanowire~\cite{Chang,Higginbotham, Krogstrup,Albrecht}, we get the topological gap $0.07$~meV, which is three times smaller than the superconducting gap in Al.

For larger energies ($\delta \lesssim |E|\lesssim |\Delta|$) we can no more consider the isolated 1D Fermi surfaces and need to apply the calculations of the topological index directly to the Hamiltonian~(\ref{two_level_system}) for $\mu = -1/2$. The resulting winding number can be either unity or zero for odd or even number of the above 1D Fermi surfaces, respectively. The switching of this winding number gives us a topological transition and a single Majorana pair at the wire edges in the topologically nontrivial phase. Thus, the topological phase transition emerges when the quasiparticle spectrum becomes inverted at zero parallel momentum $\delta \mathcal{E} = \mathcal{E}_{-1/2}(0)-\mathcal{E}_{1/2}(0)>0$. It follows from Eq.~(1) that this inversion governed by the oscillating term $\delta\mathcal{E}\propto \delta\cos\left(2k_F^cR_c\right)$ should occur periodically with the change of the number of the normal state transverse modes. An even number of the edge states generated due to the multiple 1D Fermi surfaces should, in turn, couple into the standard electronic states at nonzero energy. The remaining pair of the Majorana states localized at different edges should be robust against angular symmetry breaking perturbations provided the low-energy states are well separated from the states corresponding to $\mu\neq-1/2$ angular momenta sectors.

\begin{figure}
\centering
\includegraphics[scale = 0.65]{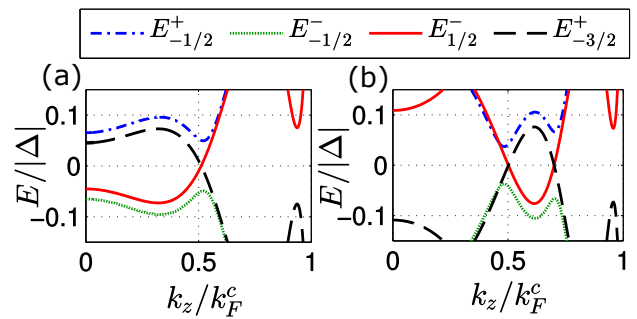}
\caption{Quasiparticle spectra in the presence of spin-orbit coupling for $\delta = 0.2|\Delta|$, $k_F^c\xi_{so} = 0.4$, $\xi_{so} = \alpha /|\Delta|$, $V_Z = 0.01|\Delta|$, $\omega_H/|\Delta| = 0.05$, $k_F^c R_c = 5$ (a), and $k_F^c R_c = 6$ (b).}
\label{Fig:spin_orbit}
\end{figure}

Before comparing our results with the results of the effective model~(5) obtained in Ref.~\cite{Lutchyn1} we should, first, emphasize an important difference between our approaches. A hollow-cylinder model (5) in Ref.~\cite{Lutchyn1} describes the modes with different angular momenta $m_J$ which are localized primarily within a thin accumulation layer near the core/shell interface. These modes provide a number of quasiparticle transport channels and this number depends on the quasiparticle energy. In Fig. 2 of Ref.~\cite{Lutchyn1} the area of the core cross-section $S$ is chosen to be of the order $(k_F^c)^{-2}$. Thus, the number of transverse modes at the Fermi level in the normal state $(k_F^c)^2S$ is of the order unity and the results presented in Fig. 2 of Ref.~\cite{Lutchyn1} refer to the situation close to a single-mode limit. On the contrary, most of our results are derived for nanowires with a flat profile of the conduction band edge in the semiconducting core (a full-cylinder model) allowing the core to host several transverse modes at the Fermi level in the normal state. Thus, we study the systems with many transport channels at the Fermi level in the normal state taking the parameter $(k_F^c)^2S$ well exceeding unity.
Comparing the expressions for the spectra~(\ref{two_level_spectrum}) from our work and (11) from Ref.~\cite{Lutchyn1}, one can see that these expressions result in qualitatively different properties both for propagating and localized quasiparticle states. Indeed, the main point in our spectrum~(\ref{two_level_spectrum}) is that in the absence of the spin-orbit coupling the first term under the square root equals zero for the set of points $k_z = \pm k_i$ while in Eq.~(11) of Ref.~\cite{Lutchyn1} the first term under the square root does not depend on the parallel momentum at all. Consequently, we arrive at qualitatively different dependencies of the decay length of zero-energy states on the spin-orbit coupling constant: the localization length is inversely proportional to the spin-orbit coupling constant in our work while in Ref.~\cite{Lutchyn1} the localization length is proportional to the spin-orbit constant.

Schematic plots of the energy levels defined by Eqs.~(\ref{two_level_spectrum}) and~(\ref{CdGM_spectrum}) are shown in Fig.~\ref{Fig:spin_orbit} where we neglect the $k_z$ dependence of the reflection amplitude $\mathcal{R}$ for simplicity. These plots clearly show that the spin-orbit interaction also results in the appearance of spin-polarized low-energy states. Indeed, we see from Fig.~\ref{Fig:spin_orbit} that the radial Rashba field does not cause the splitting for the levels $E_{1/2}^-$ and $E_{-3/2}^+$ at the Fermi level. Thus, in agreement with Eq.~\eqref{wf_spin_degenerate} the wave functions corresponding to the levels $E_{1/2}^-$ and $E_{-3/2}^+$ for energies close to the
Fermi level should have dominating spin-up and spin-down components, respectively. Certainly, various perturbations breaking the rotational symmetry in the system caused, e.g., by the presence of the disorder should lead to the mixing of the states with different angular momenta. Such mixing provides an additional coupling between the gapless states and obviously leads to the opening of additional minigaps. Comparing Fig.~\ref{Fig:spin_orbit} and Fig.~3A from Ref.~\cite{Lutchyn1}, one can see that the spin-polarized propagating modes that we discuss are absent in Ref.~\cite{Lutchyn1}. Presumably, this discrepancy can originate from the fact that the oscillations of the energy levels vs. the parallel momentum $E(k_z)$ are less pronounced when $(k_F^c)^2S$ is close to unity (this is the case for Fig.~3A of Ref.~\cite{Lutchyn1}). Alternatively, the suppression of the oscillation amplitude of energy levels can be caused by the prevailing role of the Andreev reflection from the core/shell boundary.

Considering the problem of rotational symmetry breaking one should distinguish at least two classes of perturbations: the first one originates from possible deviations of the unit vector in the Rashba interaction from the radial one and the second class arises due to the spin-independent perturbations (e.g., due to the presence of nonmagnetic impurities in the core or due to the roughness of the core/shell interface). Both these mechanisms provide the coupling between the states with different z-projections of the angular momenta and lead to the opening of additional minigaps in the spectrum which are determined by the modulus of the corresponding matrix element of the perturbation operator. In the first case the spectral minigap should be determined by the product of the spin-orbit coupling constant and the modulus of the matrix element which couples the states with different angular momenta. On the other hand, taking account of the perturbations of the second class should result in the topologically trivial gapped states.

Experimentally, the above mentioned spin polarized low-energy vortex core states can reveal themselves, e.g., in the measurements of the heat transport through the wire. Surely to probe the spin polarized modes this setup should include spin-filtering interfaces between the wire and nonsuperconducting reservoirs. Changing the temperature $T$ one can expect the appearance of the crossover between spin-polarized and unpolarized quasiparticle heat transport at $T$ comparable to the topological gap $\approx 0.8$~K, which is experimentally accessible (see, e.g., Refs.~\cite{Chandrasekhar,Linder}). One can also expect that this effect should be strongly sensitive to the parity of the number of the Fermi level crossings for a chosen (e.g., positive) $k_z$ direction. Indeed, for each level crossing there are spin-up and spin-down modes with positive and negative group velocities, respectively [see Fig.~\ref{Fig:spin_orbit}]. The presence of even number of crossings [see Fig.~\ref{Fig:spin_orbit}(b)] should lead to a partial compensation of the spin-polarized quasiparticle currents due to the different signs of group velocities at different level crossings. Changing the external magnetic field one can change the number of Fermi level crossings, which, in turn, should lead to the oscillating magnetic field dependence of the spin-polarized heat current. Note, that in the low temperature regime the heat transport through a nanowire of a finite length should be also sensitive to the contribution of the evanescent Majorana-type states.


\
\\
\

\section{Summary}\label{summary}
To sum up, our analysis of quasiparticle spectra in the vortex state of Andreev waveguides uncovers the nature of multiple topological transitions in such hybrid systems. It is shown that these transitions are driven by the interplay of the normal and Andreev reflection at the waveguide surface and accompanied by the appearance of spin-polarized propagating low-energy states. Thus, our results suggest a new way of creating and tuning multiple topologically-nontrivial states and spin-polarized states in proximized semiconducting nanowires at moderate magnetic fields.


\acknowledgements
We thank S.~Vaitiek\.{e}nas, M.-T.~Deng, C.~Marcus for stimulating discussions and I.~Khaymovich, A.~Samokhvalov, I.~Shereshevskii for valuable comments. This work was in part supported by the Russian Foundation for Basic Research under Grants No. 17-52-12044, 18-02-00390, 19-31-51019, the Foundation for the Advancement of Theoretical Physics and Mathematics ``BASIS'' Grant No. 17-11-109. In the part of numerical calculations, the work was supported by Russian Science Foundation (Grant No. 17-12-01383).

\appendix

\section{Derivation of Eq.~(1)}\label{equation_derivation}
In this section we derive Eq.~(1) in the main text keeping the notations consistent. In the limit $\Phi\lesssim \Phi_0$ for a flat confining potential profile in the SM core, the solutions of Eqs.~(6) have the following form:
\begin{eqnarray}
\Psi_{\mu, k_z}(r) = \begin{bmatrix}f^+_{\mu, k_z}(r)\\ f^-_{\mu, k_z}(r)\end{bmatrix} \ ,\\
f^+_{\mu, k_z}(r) = C^+J_{|\mu_e|}\left(\varkappa_e r\right) \ ,\\
f^-_{\mu, k_z}(r) = C^- J_{|\mu_h|}\left(\varkappa_h r\right) \ , \\
 \frac{\hbar^2\varkappa_e^2}{2m^*} = E_{\perp}^c + E - \mu_e\hbar\omega_c/2 \ ,\\
 \frac{\hbar^2\varkappa_h^2}{2m^*} = E_{\perp}^c - E + \mu_h\hbar\omega_c/2 \ ,
\end{eqnarray}
where $C^{\pm}$ are real constants, $J_{\mu}(x)$ denotes the Bessel function of the first kind of the order $\mu$, $\mu_{e,h} = (\mu \mp n/2)$, $E_{\perp}^c = (E_F^c - p_z^2/2m^*)$, and $E_F^c$ is the Fermi energy in the core. Using the WKB asymptotic form of the Bessel functions~\cite{CdGM,Bardeen} and matching the radial wave functions at the SM/SC interface using the scattering matrix defined by the amplitude reflection coefficient $\mathcal{R}$ (we choose $\mathcal{R}$ to be real), we derive the equation for the spectrum of the sub-gap states:
\begin{eqnarray}
e^{2i\alpha}\left[\mathcal{R}^2 - \mathcal{R}(e^{-2i\varphi_e} + e^{2i\varphi_h}) + e^{-2i\varphi_e + 2i\varphi_h}\right] - \\
\nonumber
 e^{-2i\varphi_e + 2i\varphi_h}\mathcal{R}^2 + \mathcal{R}(e^{-2i\varphi_e} + e^{2i\varphi_h}) - 1 = 0 \ ,\\
\nonumber
\varphi_e \approx \frac{\varkappa_e}{k_{\perp}}\left(k_{\perp}R_c + \frac{\mu_e^2}{2k_{\perp}R_c} - \frac{\pi}{2}|\mu_e|\right) - \pi/4 \ ,\\
\nonumber
\varphi_h \approx \frac{\varkappa_h}{k_{\perp}}\left(k_{\perp}R_c + \frac{\mu_h^2}{2k_{\perp}R_c} - \frac{\pi}{2}|\mu_h|\right) - \pi/4 \ .
\end{eqnarray}
Here $\alpha = \arccos\left(E/|\Delta|\right)$, $\alpha \in [0,\pi]$.  In the absence of the normal reflection $\mathcal{R} = 0$ we derive the following equation
\begin{equation}
e^{2i\alpha - 2i\varphi_e + 2i\varphi_h} - 1 = 0 \ ,
\end{equation}
which in the limit $|E|\ll E_{\perp}^c,|\Delta|$ for a singly quantized vortex has the well-known solution describing the CdGM energy levels~\cite{CdGM}:
\begin{equation}
\mathcal{E}^{(0)}_{\mu,k_z} = \omega_0\mu \ .
\end{equation}
In linear order $\propto \mathcal{R}$ we derive the correction to the CdGM spectrum~\cite{Mel'nikov,Kopnin,Kopnin_PRB_2007}:
\begin{equation}\label{CdGM_appendix}
\mathcal{E}_{\mu,k_z} = \mathcal{E}^{(0)}_{\mu,k_z}- \delta \sin(2k_{\perp}^cR_c - \pi\mu) \ ,
\end{equation}
where $\delta = 2|\Delta| \mathcal{R}$. Adding the level shifts caused by the orbital effect $\mu \omega_H/2$ and the Zeeman effect $V_Z$, we derive the Eq.~(1) in the main text. We derive the following expression describing the $\mathcal{R}^2(k_z)$ dependence for quaisiparticles at the core/shell interface in the case of rather large shell thickness $d_s\gtrsim \xi$
\begin{equation}
 \mathcal{R}^2(k_z) = \left[\frac{Z(k_z)-1}{Z(k_z)+1}\right]^2 \ , \ \ \ Z(k_z) = \frac{k_{\perp}^cm_s}{k_{\perp}^sm^*} \ .
\end{equation}
Here $k_{\perp}^s = \sqrt{(k_F^s)^2 - k_z^2}$ and $\hbar k_F^s$ is the Fermi momentum in the shell.

\section{General solutions of the BdG equations in the absence of spin-dependent interactions}\label{exact_solutions}
In this section we show the exact general solutions of the BdG equations inside the SM core~(6) neglecting the spin-dependent interactions. Considering, e.g., the equation for the electron component of the wave function and omitting the subscripts $\mu$ and $k_z$, we get
\begin{eqnarray}
\left[-\frac{d}{d\phi}\left(\phi\frac{d}{d\phi}\right) + \frac{1}{4\phi}\left(\mu_e + \phi\right)^2\right]f^+(\phi) = \\
\nonumber
 \frac{\left(E + E_{\perp}^c\right)}{\hbar \omega_c}f^+(\phi) \ .
\end{eqnarray}
Here $\phi(r) = \pi H r^2/2\Phi_0$. The regular solutions of this equation at $r = 0$ can be expressed in terms of the Kummer's function of the first kind $_1F_1(a,b,z)$~\cite{Abramowitz}. Thus, the general solutions of the BdG equations inside the SM core~(7) in the case of a constant conduction band edge have the form:
\begin{eqnarray}
f^{+}(r) = C^{+}e^{-\phi/2}\phi^{|\mu_e|/2} {_1}F_1\left(a_{\mu_e}^+,b_{\mu_e},\phi\right) \ ,\\
f^{-}(r) = C^{-}e^{-\phi/2}\phi^{|\mu_h|/2}{_1}F_1\left(a_{\mu_h}^-,b_{\mu_h},\phi\right) \ ,\\
a_{\mu}^{\gamma} = \frac{|\mu|+\gamma\mu+1}{2} - \frac{\left(E_{\perp}^c +\gamma E \right)}{\hbar\omega_c} \ ,\\
b_{\mu} = |\mu| + 1 \ ,
\end{eqnarray}
where $\gamma = \pm$. Considering the potential profile inside the SM core in the form of Eq.~(4), $U(r) = a - b r^2$, we get the following solutions for the components of the quasiparticle wave function
\begin{eqnarray}
f^+(r) = C^+e^{-\varkappa \phi/2}\phi^{|\mu_e|/2}{_1}F_1\left[a_{\mu_e}^+(\varkappa),b_{\mu_e},\varkappa \phi\right] \ ,\\
f^-(r) = C^-e^{-\varkappa \phi/2}\phi^{|\mu_h|/2}{_1}F_1\left[a_{\mu_h}^-(\varkappa),b_{\mu_h},\varkappa \phi\right] \ ,\\
a_{\mu}^{\gamma}(\varkappa) = \frac{1}{\varkappa}\biggl[\frac{\varkappa\left(|\mu| + 1\right) + \gamma \mu}{2} + \\
\nonumber
 \frac{E_{||}^c+a -\gamma E}{\hbar \omega_c}\biggl] \ ,\\
\varkappa = \sqrt{1 - 8b/m^*\omega_c^2} \ .
\end{eqnarray}
Here $E_{||}^c = p_z^2/2m^*$. The above expressions are valid for real $\varkappa$.  In the case $\varkappa =i\tilde{\varkappa}$, where $\tilde{\varkappa} = \sqrt{8b/m^*\omega_c^2 - 1}$, we derive
\begin{eqnarray}
f^+(r) = C^+ \phi^{|\mu_e|/2}\\
\nonumber
{\rm Re}\left\{e^{-i \tilde{\varkappa} \phi/2}{_1}F_1\left[a_{\mu_e}^+(i\tilde{\varkappa}),b_{\mu_e},i\tilde{\varkappa} \phi)\right]\right\} \ ,\\
f^-(r) = C^- \phi^{|\mu_h|/2}\\
\nonumber
{\rm Re}\left\{ e^{-i\tilde{\varkappa} \phi/2}{_1}F_1\left[a_{\mu_h}^-(i\tilde{\varkappa}),b_{\mu_h},i\tilde{\varkappa} \phi\right]\right\} \ .
\end{eqnarray}

\section{Derivation of the boundary conditions at the SM/SC interface}
\label{boundary_conditions_derivation}
In this section we derive the boundary conditions for the quasiparticle wave function at the SM/SC interface. Neglecting the spin-dependent interactions the solutions of the BdG equations~(2) inside the SC shell can be presented in the form of Eq.~(5), where the radial dependence is given by the following expression~\cite{CdGM,Bardeen}:
\begin{equation}
\label{Psi_ansatz}
\Psi_{\mu, k_z}(r) = g_{\mu, k_z}(r)H_{l}^{(1)}(k_{\perp}^sr) + \text{c.c.} \ .
\end{equation}
Here $g_{\mu, k_z}(r)$ is the slowly varying envelope, $H^{(1)(2)}_{l}(x)$ denote the Hankel functions of the first and the second kind of the order $l= \sqrt{\mu^2 + n^2/4}$. We use the WKB asymptotic form for the Hankel function~\cite{Bardeen}:
\begin{eqnarray}
\label{Hankel_asymptotics}
H_{l}^{(1)}(k_{\perp}^sr) \sim \exp\left(i\int_{r_t}^r\beta(r')dr'\right)\biggl/(r^2 - r_t^2)^{1/4} \ ,\\
\beta(r) = (k_{\perp}^s/r)\left(r^2 - r_t^2\right)^{1/2} \ ,
\end{eqnarray}
and $r_t = l/k_{\perp}^s$ is the turning point. Substituting Eqs.~(5) and~(\ref{Psi_ansatz}) into Eq.~(2) inside the SC shell, we get the following quasiclassical equations for the envelopes:
\begin{eqnarray}
\label{quasiclassical_eq}
-2i\check{\tau}_z\frac{dg_{\mu, k_z}}{dx} + \check{\tau}_xg_{\mu, k_z} = \Lambda g_{\mu, k_z} \ ,\\
\label{x_function}
x = \left(2m_s|\Delta|/\hbar^2k_{\perp}^s\right)\left(r^2 - r_t^2\right)^{1/2} \ ,
\end{eqnarray}
where $\Lambda  = \tilde{E}/|\Delta|$, $\tilde{E} = E - \mu \hbar V_s(R_c)/R_c$, and $V_s(R_c) = (\hbar /2m_sR_c)(\Phi/\Phi_0 - n)$. The Eq.~(\ref{quasiclassical_eq}) is valid in the case $d_s\ll R_c,\xi$. The general solution of Eq.~(\ref{quasiclassical_eq}) has the form:
\begin{eqnarray}\label{g_solution}
g_{\mu, k_z}(x) = C_1\begin{pmatrix}1\\ e^{i\varphi_s} \end{pmatrix}e^{\zeta x/2} + C_2\begin{pmatrix}1\\ e^{-i\varphi_s} \end{pmatrix}e^{-\zeta x/2} \ .
\end{eqnarray}
Here $C_{1,2}$ are the complex constants, $\varphi_s = \arccos(\tilde{E}/\zeta)$, $\varphi_s\in [0,\pi]$, and $\zeta = \sqrt{1- \Lambda^2}$. Imposing the hard-wall boundary conditions for the quasiparticle wave function at $r = R$, where $R = R_c + d_s$ is the radius of the full hybrid nanowire, we exclude a pair of complex conjugated coefficients and get
\begin{equation}
\Psi_{\mu,k_z}(r) = \begin{bmatrix}u^s_{\mu,k_z}(r)\\v^s_{\mu,k_z}(r)\end{bmatrix} \ ,
\end{equation}
where
\begin{widetext}
\begin{eqnarray}
\label{explicit_solutions}
u^s_{\mu, k_z}(r) = \bar{C}_1\biggl\{e^{\lambda/2}H_{l}^{(1)}(k_{\perp}^sr) - KH_{l}^{(2)}(k_{\perp}^sr)\left[\varkappa_2e^{\lambda^*/2} + \varkappa_1e^{-\lambda^*/2}\right]\biggl\} +\\
\nonumber
  \bar{C}_2\biggl\{e^{-\lambda/2}H_{l}^{(1)}(k_{\perp}^sr) - KH_{l}^{(2)}(k_{\perp}^sr)
\left[\varkappa_1e^{\lambda^*/2} + \varkappa_2e^{-\lambda^*/2}\right]\biggl\} \ ,\\
\nonumber
v^s_{\mu, k_z}(r) = \bar{C}_1\biggl\{\left(\Lambda + i\zeta\right)H_{l}^{(1)}(k_{\perp}^sr)e^{\lambda/2} - KH_{l}^{(2)}(k_{\perp}^sr) \left[\varkappa_2\left(\Lambda - i\zeta^*\right)e^{\lambda^*/2} + \varkappa_1\left(\Lambda + i\zeta^*\right)e^{-\lambda^*/2}\right]\biggl\} + \\
\nonumber
\bar{C}_2\biggl\{\left(\Lambda - i\zeta\right)H_{l}^{(1)}(k_{\perp}^sr)e^{-\lambda/2} - KH_{l}^{(2)}(k_{\perp}^sr)\left[\varkappa_1\left(\Lambda - i\zeta^*\right)e^{\lambda^*/2} + \varkappa_2\left(\Lambda + i\zeta^*\right)e^{-\lambda^*/2}\right]\biggl\} \ ,\\
\label{lambda_fun}
\lambda(r) = \zeta[x(r)-x_R] \ .
\end{eqnarray}
\end{widetext}
Here we introduce $K = H_{l}^{(1)}(k_{\perp}^sR)/H_{l}^{(2)}(k_{\perp}^sR)$, $\zeta' = {\rm Re}(\zeta)$, $\zeta'' = {\rm Im}(\zeta)$, $\varkappa_1 = \zeta'/\zeta^*$, $\varkappa_2 = 1- \varkappa_1$, $x_R = x(R)$, and $x(r)$ is defined by Eq.~(\ref{x_function}). One can present the solutions at the SM/SC interface in the matrix form:
\begin{eqnarray}
\label{s_solution}
\Psi_{\mu, k_z}(R_c + 0)  = \check{\mathcal{B}}\bar{C} \ ,
\end{eqnarray}
where $\bar{C} = [\bar{C}_1,\bar{C}_2]^{T}$ and the matrix elements of $\check{\mathcal{B}}$ are easily written using Eq.~(\ref{explicit_solutions}).
Provided that $k_{\perp}^sR_c \gg 1$ and $|\mu|\ll k_{\perp}^sR_c$, let us present the expressions for the derivative of the solutions at the interface:
\begin{equation}
\label{s_solution_derivative}
\frac{d\Psi_{\mu, k_z}}{dr}\biggl|_{r = R_c + 0} = \check{\mathcal{A}}\bar{C} \ ,
\end{equation}
where
\begin{eqnarray}
\mathcal{A}_{11} = i\beta_ce^{\lambda_c/2}H_{l}^{(1)} +\\
\nonumber
 i\beta_cKH_{l}^{(2)}\left(\varkappa_2e^{\lambda_c^*/2} + \varkappa_1e^{-\lambda_c^*/2}\right) \ ,\\
\nonumber
\mathcal{A}_{22} = i\beta_c(\Lambda - i\zeta)e^{-\lambda_c/2}H_{l}^{(1)} + i\beta_cKH_{l}^{(2)} \cdot \\
\nonumber
\left[\varkappa_1(\Lambda - i\zeta^*)e^{\lambda_c^*/2} + \varkappa_2(\Lambda + i\zeta^*)e^{-\lambda_c^*/2}\right] \ ,\\
\nonumber
\mathcal{A}_{12} = \mathcal{A}_{11}\left(\zeta \to -\zeta\right) \ ,\\
\nonumber
\mathcal{A}_{21} = \mathcal{A}_{22}\left(\zeta \to -\zeta\right) \ .
\end{eqnarray}
Here $\lambda_c = \lambda(R_c)$, $\lambda(r)$ is defined by Eq.~(\ref{lambda_fun}), $\beta_c = (k_{\perp}^s/R_c)\sqrt{R_c^2 - r_t^2}$, and here we omit the arguments of the Hankel functions $k_{\perp}^sR_c$ for brevity. Finally, combining Eq.~(\ref{s_solution}),~Eq.~(\ref{s_solution_derivative})
\begin{equation}
\frac{d\Psi_{\mu,k_z}}{dr}\biggl|_{r = R_c + 0} = \check{\mathcal{A}}\check{\mathcal{B}}^{-1}\Psi_{\mu,k_z}(R_c)
\end{equation}
 and imposing
\begin{eqnarray}
\label{homogeneous_bc_core}
\Psi_{\mu, k_z}(R_c - 0) = \Psi_{\mu, k_z}(R_c + 0) \ ,\\
\nonumber
\frac{d\Psi_{\mu, k_z}}{dr}\biggl|_{r = R_c - 0} = \left(\frac{m^*}{m_s}\right)\frac{d\Psi_{\mu, k_z}}{dr} \biggl|_{r = R_c + 0} \ ,
\end{eqnarray}
we derive the boundary conditions for the wave functions at the SM/SC interface
\begin{eqnarray}
\label{boundary_conditions}
\frac{d \Psi_{\mu, k_z}}{dr} \biggl|_{r = R_c - 0} = \\
\nonumber
\frac{1}{\zeta}\begin{pmatrix} \zeta M' - \Lambda M''&M''\\ - M''&\zeta M'+\Lambda M''\end{pmatrix}\Psi_{\mu, k_z}(R_c - 0) \ ,\\
\label{M_expression}
M = -\frac{m^*}{m_s}k_{\perp}^s\cot\left(k_{\perp}^sd_s - i\zeta\frac{m_s|\Delta| d_s}{\hbar^2 k_{\perp}^s}\right) \ .
\end{eqnarray}
Here $M' = {\rm Re}(M)$ and $M'' = {\rm Im}(M)$.
It follows from the~(\ref{M_expression}) that for rather thin shell $d_s\lesssim \hbar^2k_{\perp}^s/m_s|\Delta|$ there appear mesoscopic oscillations of the energy levels caused by the presence of additional normal scattering for quasiparticles at the external shell boundary. In the limit $d_s\gtrsim \xi$ we get
\begin{equation}
\label{M_expression_Andreev}
M = ik_{\perp}^s m^*/m_s \ .
\end{equation}

\section{Derivation of Eq.~(8)}\label{spin_orbit_splitting_derivation}
In this section we derive Eq.~(8) in the main text. As mentioned in the Ref.~\cite{Lutchyn1} in the presence of the spin-orbit interaction $\hat{H}_{so} = (\alpha/\hbar)\mathbf{e}_{r}[\hat{\mathbf{\sigma}},(\mathbf{p} + eA_{\varphi}\mathbf{e}_{\varphi})]$
the solutions of the BdG equations~(2) can be presented in the form:
\begin{equation}\label{general_qp_wave_function}
\Psi(\mathbf{r}) = e^{ik_zz}e^{i\left[\mu + (1-\hat{\sigma}_z)/2 - \check{\tau}_zn/2\right]\varphi}\Psi_{\mu, k_z}(r) \ .
\end{equation}
Substituting Eq.~(\ref{general_qp_wave_function}) into Eq.~(2) taking into account the spin-orbit coupling inside the SM core, one gets the following radial eigenvalue problem in the case of a flat conduction band edge in the core:
\begin{eqnarray}\label{general_radial_eigenvalue_problem}
\check{\mathcal{L}}\Psi_{\mu, k_z}(r) = E\Psi_{\mu, k_z}(r) \ ,\\
\nonumber
\check{\mathcal{L}} = \left( - \frac{\hbar^2}{2m^*}\frac{1}{r}\frac{d}{d r} r \frac{d}{d r} - E_{\perp}^c\right)\check{\tau}_z + V_Z\hat{\sigma}_z +\\
\nonumber
\frac{\hbar^2}{2m^*r^2}\left(\mu + \hat{\Pi}_z^- - \frac{n}{2}\check{\tau}_z + \frac{\pi H r^2}{2\Phi_0}\check{\tau}_z\right)^2\check{\tau}_z - \\
\nonumber
\frac{\alpha}{r}\hat{\sigma}_z\check{\tau}_z\left(\mu + \hat{\Pi}_z^- - \frac{n}{2}\check{\tau}_z + \frac{\pi H r^2}{2\Phi_0}\check{\tau}_z\right) +\alpha p_z\check{\tau}_z\hat{\sigma}_y \ ,
\end{eqnarray}
where $\hat{\Pi}_z^- = (1-\hat{\sigma}_z)/2$. Substituting Eq.~(7) into Eq.~(\ref{general_radial_eigenvalue_problem}) and neglecting the spin-orbit induced diagonal corrections to the CdGM levels, we derive Eq.~(8). Within the low energy effective model described by the Eqs.~(\ref{two_level_system}) for $\mu = -1/2$ and~(\ref{linearized_spectrum}), every pair of the Fermi level crossings at $k = \pm k_i$ produces zero energy evanescent modes at the nanowire edges. Considering a semi-infinte wire $z\in [0,+\infty)$ we derive the following explicit expressions for zero-enery states
\begin{eqnarray}
 \Psi(\mathbf{r}) = C\sin\left(\beta_iz\right)e^{-\gamma_iz}\begin{bmatrix}u_{\mu,k_i}(r)e^{-i\varphi}\\ v_{\mu,k_i}(r)\\ {\rm sgn}(v_g^i)u_{\mu+1,k_i}(r)\\ {\rm sgn}(v_g^i)v_{\mu + 1,k_i}(r)e^{i\varphi}\end{bmatrix} \ ,\\
 \nonumber
 \beta_i = \frac{k_i}{1 + (\alpha/v_g^i)^2} \ , \ \ \  \gamma_i = \frac{\alpha k_i}{|v_g^i|[1 + (\alpha/v_g^i)^2]} \ ,
\end{eqnarray}
where we choose $k_i > 0$ and $C$ is a normalization constant.


\begin{references}
\bibitem{n=1_footnote}
The winding number of the superconducting order parameter is considered to be $n = 1$.
\bibitem{Self_cons_Delta_footnote} More rigorous expression for $\Delta(\mathbf{r})$ profiles should be determined from the solution of the self-consistency equation for the pair potential in the shell~\cite{Virtanen,Tanaka,Peeters}.
\bibitem{Self-consist_U_footnote}
More rigorous approach of taking into account the confining potential $U(r)$ should rely on self-consistent calculations of the band edge and charge density profiles~\cite{Demarina,Antipov,Mikkelsen,Woods2}.
\end{references}
\end{document}